\documentclass[a4]{revtex4}

\usepackage{graphicx}
\usepackage{amsmath}
\begin{document}

\title{Accurate multiple time step in biased molecular simulations}

\author{Marco Jacopo Ferrarotti}

\author{Sandro Bottaro}

\author{Andrea P{\'e}rez-Villa}

\author{Giovanni Bussi}
\email{bussi@sissa.it}

\affiliation{Scuola Internazionale Superiore di Studi Avanzati (SISSA), via Bonomea 265, 34136 Trieste, Italy}

\begin{abstract}
Many recently introduced enhanced sampling techniques are based on
biasing coarse descriptors (collective variables) of a molecular system
on the fly. Sometimes the calculation of such collective variables is
expensive and becomes a bottleneck in molecular dynamics simulations.
An algorithm to treat smooth biasing forces within a multiple time step framework
is here discussed. The implementation is simple and allows a speed
up when expensive collective variables are employed. The gain can
be substantial when using massively parallel or GPU-based molecular
dynamics software. Moreover, a theoretical framework to assess the
sampling accuracy is introduced, which can be used to assess the
choice of the integration time step in both single and multiple time
step biased simulations.
\end{abstract}

\maketitle

\section{Introduction}
Molecular simulations allow the dynamics of a system to be followed
at atomistic resolution and thus can be used as virtual microscopes
to investigate chemical reactions and conformational transitions in
molecular systems \cite{alle-tild87book,frenk-smit02book,citeulike:5961095}.
The time scale that can be simulated with current computers and algorithms
is on the order of a microsecond for a system of a few tens of thousands of atoms
modeled with a typical atomistic empirical
force field.
Even though recently 
developed \emph{ad hoc }hardware allowed for
a sudden jump in this scale
\cite{dror2011anton}, many relevant processes are still out
of range for atomistic molecular dynamics (MD).
A lot of effort has been made in
the last decades to tackle or, at least, to alleviate this issue
by means of modified algorithms known as enhanced sampling techniques.
Among these, a class of methods is based on the idea of biasing a
few, carefully selected collective degrees of freedom, known as collective
variables (CVs) \cite{torri-valle77jcp,darv-poho01jcp,laio-parr02pnas,mara-vand06cpl,mars+06jpcb,bard+08prl,abrams2008efficient,laio-gerv08rpp,barducci2011metadynamics,abrams2013enhanced,dellago2013computing}.
Biased enhanced-sampling simulations require these descriptors to
be computed on the fly at every simulation time step. The cost of
computing the CVs can in principle slow down the calculation. This
is particularly true for expensive CVs, such as Steinhardt order parameters
\cite{stei84prb,trudu2006freezing}, path CVs \cite{bran+07jcp},
template-based CVs \cite{Pietrucci:2009df}, property maps \cite{Spiwok:2011ce},
social permutation-invariant coordinates \cite{pietrucci2011graph},
sketch maps \cite{tribello2012using} and Debye-H\"uckel energy \cite{do+13jctc},
just to mention a few. The calculation of the CVs and of the resulting
biasing forces can be done in dedicated routines of MD software or can be
implemented in external plugins \cite{bono+09cpc,fiorin2013using,tribello2013plumed2}.
It is very difficult to optimize the calculation of the CVs for several
reasons. First, they sometime involve long-range exchange of information
that makes them difficult to be compatible with domain-decomposition
algorithms. Second, because their definition is highly system dependent,
routines implementing these schemes should be flexible, thus paying
an unavoidable price in terms of efficiency. On the opposite
side, the calculation of physical force fields is becoming faster
and faster thanks to the use of massively parallel and optimized software
\cite{plim95jcp,spoe+05jcc,NAMD} and hardware \cite{dror2011anton}.
The overall consequence of this trend is that the calculation of CVs
can be expected to become a relevant bottleneck in enhanced sampling
simulations, particularly in the context of massively parallel or
GPU-accelerated MD.

We here introduce a scheme based on multiple time step integration
(MTS) in its reference system propagator algorithm (RESPA) formulation
\cite{tuckerman:1990,sext-wein92npb}. More precisely, MTS is used
here to split the integration of biasing forces and physical forces.
The underlying principle is that often the former have a smoother
dependence on atomic positions, thus change at a lower rate and can
be integrated with a larger time step. This splitting allows for a
substantial increase in the computational efficiency of biased sampling
methods. Furthermore, we introduce a theoretical framework so as to
assess the unavoidable time-step discretization errors. This is done
by extending the concept of \emph{effective energy} \cite{buss+07jcp,buss-parr07pre}
so as to take into account the effect of biasing forces separately.
This scheme allows artifacts that could be hidden when observing total-energy conservation to be highlighted. In the Method Section we introduce
the methodology, first reviewing the MTS integration then introducing
the scheme to assess sampling accuracy. We then present numerical
tests of our algorithm on a one-dimensional model, on alanine dipeptide
and on a large protein/RNA complex, using CVs of increasing complexity.
In the latter case we show that the use of MTS can greatly improve
the performance without introducing significant errors.

\section{Methods}
\subsection{Multiple time step}
We consider here a system of $N_{at}$ atoms evolving accordingly
to biased Hamilton equations in the form:
\begin{align*}
\dot{\mathbf{q}}_{i} & =\frac{\mathbf{p}_{i}}{m_{i}}\\
\dot{\mathbf{p}}_{i} & =-\frac{\partial U}{\partial\mathbf{q}_{i}}-\frac{\partial V}{\partial\mathbf{q}_{i}}=-\frac{\partial U}{\partial\mathbf{q}_{i}}-\sum_{\alpha=1}^{N_{cv}}\frac{\partial V}{\partial s_{\alpha}}\frac{\partial s_{\alpha}}{\partial\mathbf{q}_{i}}\,.
\end{align*}
Here $\mathbf{q}_{i}$, $\mathbf{p}_{i}$, and $m_{i}$ are respectively
 coordinates,  momenta, and mass of $i$-th atom; $U(q)$ is the physical
potential, typically obtained evaluating an empirical force field
(see, e.g., Ref.~\cite{corn+95jacs}); $V(s)$ is a biasing potential, which
depends on the atomic positions only through $N_{cv}$ collective
variables $s_{\alpha}(q)$.

The choice of the time step $\Delta t$ for the integration of the
equations of motion is dictated by the fastest frequencies of the system,
which depend on the atomic masses and on the functional form of $U$.
In reversible MTS integration forces are typically split in a fast-changing
part and slow-changing part, so that different integration time steps
can be used for the two parts \cite{tuckerman:1990,sext-wein92npb}.
Ideally, the slow-changing part is the most computationally expensive
one.

We remark here that CVs are often smoother functions of atomic coordinates
than the stiffest force-field terms. As an example, CVs used for isomerization
processes such as torsional angles are typically slower when compared
with bond fluctuations. Biasing forces are thus good candidates for
the time-reversible MTS scheme itemized below: 

\begin{enumerate}
\item Integrate momenta according to biasing force $-\frac{\partial V}{\partial\mathbf{q}_{i}}$
for a single time step of length $\frac{n\Delta t}{2}$.
\item Integrate momenta and positions according to the physical forces $-\frac{\partial U}{\partial\mathbf{q}_{i}}$
for $n$ times, with time step $\Delta t$.
\item Integrate momenta according to biasing force $-\frac{\partial V}{\partial\mathbf{q}_{i}}$
for a single time step of length $\frac{n\Delta t}{2}$.
\end{enumerate}
Here $\Delta t$ is the inner time step and $n$ is the order of the
MTS scheme, so that biasing forces and, as a consequence, CVs are
only evaluated with a time step $n\Delta t$. This algorithm can be
straightforwardly integrated in existing codes. Indeed, it is sufficient
to apply the biasing forces every $n$ steps scaling them by a factor
$n$ so as to take correctly into account the transferred momentum.
In the intermediate $(n-1)$ steps, CVs do not need to be computed.
The algorithm is by construction symplectic and reversible, as long
as the underlying integrator for the Hamilton equations is symplectic
and reversible. 

We note that a too large outer time step $n\Delta t$ can lead
to the well-known occurrence of resonances \cite{bies+93jcp}
and thus to
significant systematic errors.
Several possible solutions have been recently proposed to
tackle this issue.\cite{morr+11jcp,leimkuhler2013stochastic}
These recent improvements could be easily coupled with the presented algorithm
and might decrease the systematic errors.
We also observe that so-called multigrators \cite{pred+jcp13} do not suffer from resonances
because they are based on splitting in two equations of motion both
having the same stationary distribution and thus can be used with
arbitrarily large $n$. Such an approach is not suitable in the context
discussed here. However, we notice that in typical situations even
small values of $n$ are already enough to largely decrease the effective
computational cost of CV calculation.
Finally, we observe that MTS
has been also employed in Ref.~\cite{abrams2008efficient} with
a different purpose, i.e. to allow for a large time step for the physical
force-field to be combined with a short time step for the integration
of the CVs.

\subsection{Assessment of sampling error}
In order to provide results free from systematic errors, enhanced
sampling methods should be paired with integration of the equations
of motion able to correctly sample the target statistical ensemble,
which is typically the isothermal or the isobaric one.
In general, the errors in the stationary distribution depends
in a non-trivial way on the integration algorithm and on
the details of the underlying potential (see, e.g., Ref. \cite{leim-matt13amrx})
so that it is very useful to have error indicators that can be used
in practical simulations.
For example in microcanonical
simulations it is customary to check energy conservation. A similar
procedure can be used e.g. with the Nos\'e-Hoover thermostat \cite{nose84jcp,hoov85pra},
where a conserved energy can be defined, and has been generalized
to stochastic simulations through the definition of an \emph{effective
energy} drift\cite{buss+07jcp,buss-parr07pre}.
The effective energy represents the total energy of the extended system
(physical system plus thermostat), and its drift has been interpreted
as the work performed by the integrator on the system \cite{sivak2013using}.
We here extend this idea to MTS biased simulation. Furthermore, we
show how it is possible to compute separately the portion of energy
drift due to biasing forces alone.
This makes it possible to assess the accuracy with which
biasing forces are integrated, thus providing a safe
framework that can be used to choose $n$ in practical cases. We also
notice that this approach to assess integration accuracy is not limited
to MTS calculations, but can be used also in single time step biased
calculations ($n=1$) to verify that the biasing forces are compatible
with the chosen $\Delta t$.

\subsubsection{Effective Energy Derivation and Analysis\label{sub:Effective-Energy-Derivation}}
Effective energy drift\cite{buss+07jcp,buss-parr07pre} was introduced  
as a direct measure of the detailed balance violation:

\begin{equation}\label{eq:delta-h-tilde}
\Delta\tilde{H}=-k_{B}T\log\left(\frac{M\left(x^{*}(t)\leftarrow x^{*}(t+\Delta t)\right)}{M\left(x(t+\Delta t)\leftarrow x(t)\right)}\right)+H\left(x(t+\Delta t)\right)-H\left(x(t)\right)
\end{equation}
Here $H$ is the Hamiltonian, $\tilde{H}$ the effective energy, $x(t)$
denotes the state of the system (positions and momenta) at time $t$,
$k_{B}T$ the thermal energy, and $M(x'\leftarrow x)$ the probability
of choosing $x'$ as coordinates at the next step provided the system
is in $x$ at the present step. We here use the notation $x^{*}$
to denote the reversal of the velocities, so that strictly speaking
the drift measures \emph{generalized} detailed balance violation \cite{gard03book}.
The effective energy is then conventionally initialized to zero and
computed as a sum over time

\[
\tilde{H}(t)=\sum_{t'<t}\Delta\tilde{H}(t)
\]

Assuming integration is performed using velocity Verlet \cite{frenk-smit02book,citeulike:5961095},
Eq. \ref{eq:delta-h-tilde} can be rewritten as \cite{buss-parr07pre}
\begin{equation}\label{eq:h-drift}
\Delta\tilde{H}(t)=\sum_{i=1}^{N_{at}}\Delta\mathbf{q}_{i}\cdot\left(\frac{\mathbf{f}^{(tot)}_{i}(t)+\mathbf{f}^{(tot)}_{i}(t+\Delta t)}{2}\right)+\Delta U + \Delta V+\frac{\Delta t^{2}}{8m}\sum_{i=1}^{N_{at}}\Delta\left|\mathbf{f}^{(tot)}_{i}\right|^{2},
\end{equation}
where $\mathbf{f}_{i}^{(tot)}=-\frac{\partial U}{\partial \mathbf{q}_i}-\frac{\partial V}{\partial \mathbf{q}_i}$
is the total force (physical plus bias) acting on \emph{i-}th atom. Here $\Delta U=U(t+\Delta t)-U(t)$ represents the change
in the physical potential, and $\Delta V=V(t+\Delta t)-V(t)$ the change in the bias potential. We remark that
this expression is exact, except for numerical roundoff, when equations of motions are integrated using velocity Verlet and when
system is thermostated using a scheme that exactly preserves detailed balance.
Any thermostating procedure could then be used, 
provided it is implemented
on top of velocity Verlet (see, e.g., Appendix E in Ref.\cite{frenk-smit02book}),
and provided the integration of the thermostat is accurate.

One can make some considerations about Eq.~\ref{eq:h-drift}:

\begin{enumerate}
\item The first term represents a numerical path integral of the forces
with trapezoidal rule. Therefore, in the limit of small $\Delta t$,
its value tends to compensate the second term.
\item The second and third terms are exact differentials, therefore they
contribute to fluctuations of the effective energy and not to its
drift. In the limit of small $\Delta t$, the third term can be neglected
with respect to the second one.
\item No assumption has been made about the protocol used to evaluate such
forces, which in principle could differ from the actual gradient of
the potential energy. For instance, if $f$ is computed as an approximate
estimate of the actual force, the effective energy drift will still
measure exactly the violation of detailed balance for the correct
Hamiltonian. Thus, the effective energy drift will be smaller if better
estimates of the forces are used.
\item The first and second terms are linear in the force and in the potential
energy, so that they can be split in a contribution coming from the
physical forces and a contribution coming from the biasing forces.
\end{enumerate}

Point 3. is particularly important in this context. Indeed, one
can consider the MTS algorithm as the application of artificially
modified forces. Namely, biasing forces are neglected on some steps and
augmented by a factor $n$ on other steps. The forces that should
be used in evaluating Eq.~\ref{eq:h-drift} are thus the artificially
modified forces, whereas the potential $U+V$ should be the correct
one. Provided the artificially modified forces are used in a symplectic
integrator such as velocity Verlet, the effective energy drift
can be used to
check detailed-balance violations and, thus, to detect sampling errors.

Based on these considerations, we define an approximated bias effective energy
drift as 
\begin{equation}\label{eq:bias-drift}
\Delta\tilde{H}^{(b)}(t)=\sum_{i=1}^{N_{at}}\Delta\mathbf{q}_{i}\cdot\left(\frac{\mathbf{f}_{i}^{(b)}(t)+\mathbf{f}_{i}^{(b)}(t+\Delta t)}{2}\right)+\Delta V.
\end{equation}
Here $V$ denotes the biasing potential and $\mathbf{f}_{i}^{(b)}=-\frac{\partial V}{\partial\mathbf{q}_{i}}$
the biasing force on $i$-th atom.
We notice that here only the change in the biasing potential $\Delta V = V(t+\Delta t) - V(t)$ is considered.
The bias effective energy drift
is therefore equal to the difference between the actual increment
of bias potential and the estimate of that increment through a numerical
path integration of the biasing force. We underline here that in a
MTS framework $\Delta t$ is the inner time step and not the one used
to compute the bias potential. However, the biasing force $\mathbf{f}_{i}^{(b)}$
is only contributing at steps that are multiple of the MTS factor
$n$. In other words, in Eq. \ref{eq:bias-drift} the biasing force is
scaled by a factor $n$ and the increment in position is considered for the inner step. If the coordinates on which the biasing force
is acting were evolved at constant velocity in the inner integration
step (i.e. if the physical potential did not affect them) then this
expression would be equal to a numerical path integral of the forces.
Discrepancies from this behavior are caused by the unavoidable interplay
between the CV dynamics and the microscopic one and can be ultimately
ascribed to resonance effects.
The goal of our procedure is indeed to highlight these effects
and quantify them.

We remark here that, whereas Eq.~\ref{eq:h-drift} is exact and can be derived just 
enforcing detailed balance, the expression in Eq.~\ref{eq:bias-drift}
is approximate.
A few comments can be made comparing these two expressions.
First, we notice that the last term of Eq.~\ref{eq:h-drift} has been neglected in
Eq.~\ref{eq:bias-drift}. More precisely, since our goal is to track
any violation of detailed balance  that is due to the bias, an additional
term
\begin{equation}\label{eq:alpha}
\alpha(t)=\frac{\Delta t^{2}}{8m}\sum_{i=1}^{N_{at}}\left|\mathbf{f}_{i}^{(b)}(t)\right|^{2}+\frac{\Delta t^{2}}{4m}\sum_{i=1}^{N_{at}}\mathbf{f}_{i}^{(b)}(t)\cdot\mathbf{f}_{i}(t)
\end{equation}
should have been included in Eq. \ref{eq:bias-drift}. As we will show in the numerical examples,
this term is negligible in practical situations. Moreover, by construction
it does not contribute to the overall effective energy drift, but only to its
fluctuations. As a consequence, the calculation of $\tilde{H}^{(b)}$
only requires the knowledge of the biasing force. This allows in principle
to implement it efficiently also in external plugins \cite{bono+09cpc,fiorin2013using,tribello2013plumed2}
that do not have access to physical forces computed in the MD software.
Moreover, we observe that Eq.~\ref{eq:bias-drift} can be straightforwardly
parallelized, also when domain decomposition algorithms are employed,
and requires the reduction of a single scalar number thus with a minimal
impact on the performance.

Once the change in effective energy per time step has been computed with
Eq.~\ref{eq:bias-drift}, the effective energy can be
computed as
\[
\tilde{H}^{(b)}(t)=\sum_{t'<t}\Delta\tilde{H}^{(b)}(t')
\]

\section{Numerical examples}
\subsection{One-dimensional double-well potential}

\begin{figure}
\includegraphics[clip,scale=0.3]{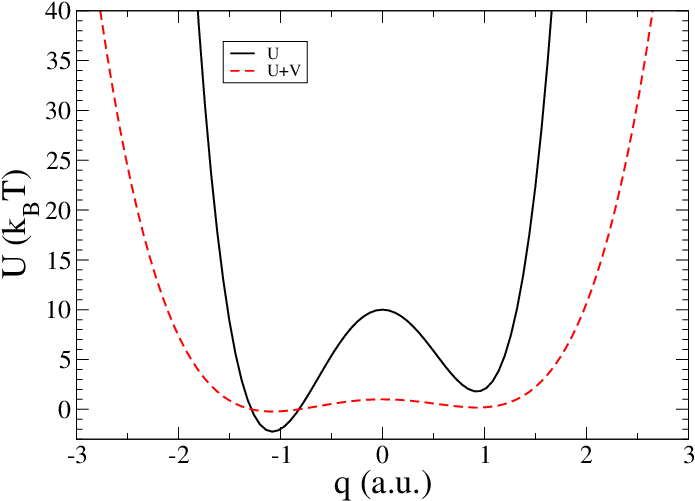}

\caption{\label{fig:potential}The model 1D potential is shown. Unbiased potential
$U$ (solid) and total potential $U+V$ (dashed), as defined in Eqs.
\ref{eq:1d-u} and \ref{eq:1d-v}. Notice that the barriers is decreased
by a factor 10 in the biased potential. }
\end{figure}
As a first example we discuss a simple one dimensional toy model of
a single particle confined into a double well potential. All quantities
are here expressed in arbitrary units, and we assume the thermal energy
$k_{B}T=1$ and particle mass $m=1$. We consider a potential energy:

\begin{equation}
U\left(q\right)=A\,\left(1-q^{2}\right)^{2}+B\, q^{3},\label{eq:1d-u}
\end{equation}
where $A=10$ and $B=2$. As shown in Fig. \ref{fig:potential} such
potential has two stable minima which are separated by a barrier of
several $k_{B}T$. Transitions are accelerated by means of an umbrella
potential \cite{torri-valle77jcp} in the form 
\begin{equation}
V\left(q\right)=-0.9\, U\left(q\right).\label{eq:1d-v}
\end{equation}
Using a negative fraction of the potential energy as bias potential
is convenient because, as it is evident from Fig. \ref{fig:potential},
it lowers the barriers keeping the system confined into the interesting
regions of phase space. Moreover, this choice is consistent with the
expected bias potential obtained in the long-time limit in a well-tempered metadynamics
simulation \cite{bard+08prl}.

\begin{figure}
\includegraphics[clip,scale=0.3]{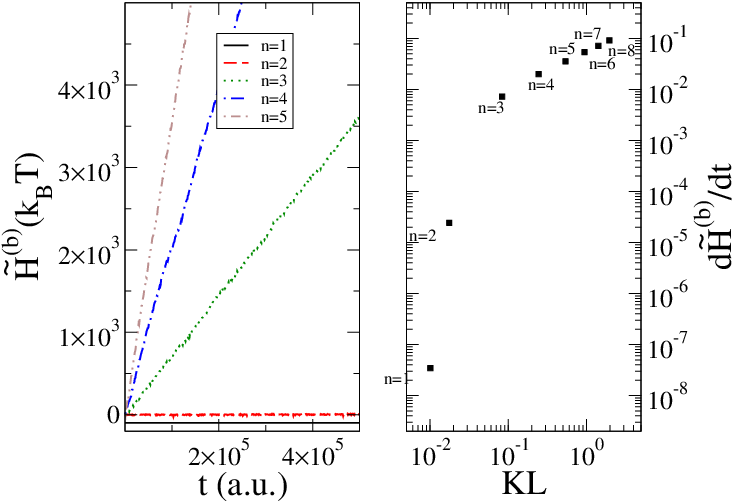}
\caption{\label{fig:drift}Bias effective energy drift as a function of time (left panel) for different
values of the multiple time step order $n$, as indicated. Notice
that the line corresponding to $n=1$ has been shifted downward by
100 $k_{B}T$ for clarity. Time derivative of bias effective energy as
a function of Kullback\textendash{}Leibler divergence $D_{KL}$ (right
panel) for $n$ ranging from 1 to 8, as indicated. Both the drift
and the $D_{KL}$ are growing monotonically with $n$.}
\end{figure}

The system is then evolved according to a Langevin equation: 

\[
m\ddot{q}=-\frac{\partial U}{\partial q}-\frac{\partial V}{\partial q}-\gamma m \dot{q}+\sqrt{2 \gamma mk_{B}T}\eta(t), 
\]

with friction $\gamma=1$. Equations are integrated with the scheme
proposed in Ref.~\cite{buss-parr07pre}, using an inner time step
$\Delta t=0.01$. MTS is implemented by scaling the biasing force
$-\frac{\partial V}{\partial q}$ by a factor $n$ and by applying
it every $n$-th step. 

The bias effective energy drift is then computed and shown in Fig.
\ref{fig:drift} for $n=1$, 2, 3, 4, and 5. Notice that the bias effective
energy is clearly drifting for $n>2$, and that its drift is growing
with $n$. This is in agreement with the intuitive observation that
a larger time step would make the path integral of the forces more
and more different from the corresponding potential, as it can be
seen in Eq. \ref{eq:bias-drift}. We observe that the actual value
of the drift also depends on the Langevin thermostat and that the
drift could be reduced by decreasing the friction $\gamma$ or by
integrating the thermostat in the outer time step (see Figs.~S1, S2, and S3). However this 1D model is meant to mimic a dimensional reduction
of a finer grain model. Thus friction should not be considered
as a tunable parameter but as an actual part of the model which depends
on how coupled is the biased CV with the other microscopic degrees
of freedom.

We also compute the difference between
the observed distribution and the theoretical one
using the Kullback-Leibler divergence
$D_{KL}=\sum_{i}\log\left(\frac{p_{i}}{h_{i}}\right)p_{i}$,
where $p_{i}$ is the theoretical distribution and $h_{i}$ is the
empirical histogram, both computed with a bin size of 0.01. In Fig.
\ref{fig:drift} we show the relationship between the slope of the
effective energy drift and the Kullback-Leibler divergence. The results
clearly show that the effective-energy drift is a good proxy for the sampling error,
and can be used in situations where a reference result does not exist
so as to simplify the choice of $n$ and to allow errors to be detected.

\begin{figure}
\includegraphics[clip,scale=0.3]{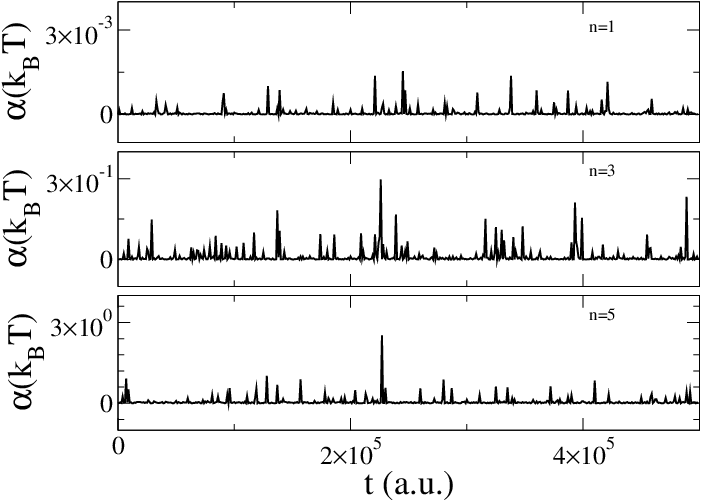}
\caption{\label{fig:alpha_term}In the three panels neglected terms $\alpha$ (see Eq. \ref{eq:alpha}) for the
bias effective energy are shown as
a function of time for selected choices of the multiple time step order
$n=1,3,5$, as indicated.
Although for $n=5$ there are spikes on the order of a few $k_BT$,
this term does not present any significant drift.
}
\end{figure}

To verify that the neglect of terms related to forces defined in Eq.~\ref{eq:alpha}
is not affecting the overall effective energy drift, we compute it explicitly.
Figure~\ref{fig:alpha_term} shows that such a force term only contributes 
to fluctuations and not to the overall drift.

The tests on this one-dimensional potential were performed using Langevin dynamics.
The idea of defining an effective energy drift in a stochastic equation of motion is
relatively new.\cite{buss+07jcp,buss-parr07pre}
However, the theoretical considerations discussed in this paper 
apply to any thermostating procedure, including deterministic ones.
As it can be appreciated in Fig.~S4, the drift
computed using Eq.~\ref{eq:bias-drift} is close to the one obtained by monitoring the total
energy conservation with a conventional Nos\'e-Hoover thermostat.\cite{nose84jcp,hoov85pra}

\subsection{Dihedral angles in alanine dipeptide}
\begin{figure}
\includegraphics[clip,scale=0.9]{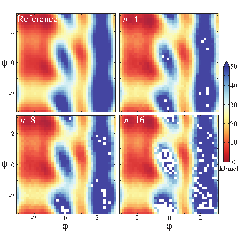}
\caption{\label{fig:fig3}Free energy surfaces of alanine dipeptide projected
onto the $\phi,\psi$ dihedral angles obtained from umbrella sampling
simulations. The reference FES (top left panel) is compared with results
obtained using different stride values of the multiple time step order $n$,
as labeled.}
\end{figure}

As a second test, we study the free energy surface (FES) of the alanine
dipeptide with umbrella sampling simulations, using different values for the multiple time step order $n$. Alanine dipeptide in water is a
standard model system for many theoretical and computational studies
of biomolecules \cite{rossky1979model}, and is known to exhibit multiple
minima on the FES computed as a function of the $\phi,\psi$ torsional
angles (see Fig. \ref{fig:fig3}). The terminally blocked alanine
peptide (sequence Ace-Ala-Nme) was solvated with 542 TIP3P \cite{jorg+83jcp}
water molecules in a rhombic dodecahedral box, whose dimensions were
chosen to ensure a minimum distance to the box boundaries of 10 \AA.
Molecular dynamics simulations were performed with the AMBER99SB \cite{hornak2006comparison}
force field using Gromacs 4.6 \cite{hess+08jctc} in combination with
an in house version of PLUMED 2 \cite{tribello2013plumed2}. Long
range electrostatic forces were treated using particle-mesh Ewald
summation \cite{dard+93jcp}, and the equations of motion were integrated
with time step $\Delta t=$2 fs. All simulations were performed in
the canonical ensemble ($T=300$K) with stochastic velocity rescaling
\cite{buss-parr07pre}, and bond lengths were constrained using the
LINCS algorithm \cite{hess2008jctc}.

\begin{figure}
\includegraphics{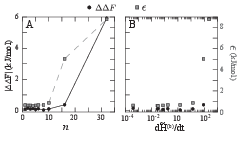}
\caption{\label{fig:fig4}Two different estimates of the sampling error in
umbrella sampling simulations of alanine dipeptide, namely free-energy
difference between two regions ($\Delta\Delta F$, circles) and average
error ($\epsilon$, squares). Left: Sampling error vs multiple time
step order. Right: Sampling error vs time derivative of bias effective energy drift. }
\end{figure}

We first performed a 100 ns well-tempered metadynamics simulation
\cite{laio-parr02pnas,bard+08prl} using the $\phi,\psi$ angles as
collective variables. The Gaussian width was set to 0.35 rad, and
the deposition interval was 3.2 ps with a starting Gaussian height
of 0.4 kJ/mol and a bias factor of 10, corresponding to an enhancement
in the CV temperature of $\Delta T=2700$ K. We then used the obtained
bias potential to perform an umbrella sampling (US) refinement as
in Ref.\cite{babi+06jcp}. A reference free energy landscape was
obtained by running a 500ns US simulation with single time step ($n=1$).
We then performed a set of 100 ns US simulations using different multiple
time step orders $n$, and compared the results with the reference
simulation. As for the 1D double-well case, the accuracy on the reconstructed
free energy surface decreases for increasing values of $n$ (Fig.~\ref{fig:fig3}).
The discrepancy is noticeable around the free energy maxima, but much
less pronounced in the relevant low-free-energy regions. The agreement
between the reference and the MTS simulation with stride $n$ was
quantified by calculating the deviation 

\[
\Delta\Delta F=\Delta F_{REF}-\Delta F_{MTS,n}
\]

 where $\Delta F_{REF}$ and $\Delta F_{MTS,n}$ are the free energy
differences between the region $\phi\in[-\pi,-0.4]$ and the region
$\phi\in[0.4,1.6]$ for the reference and the MTS simulation, respectively.
As a measure of the error, we additionally computed the deviation
from the reference free-energy landscape as 

\[
\epsilon=\left(\frac{1}{A}\int_{\Gamma}\left[F_{REF}(\phi,\psi)-F_{MTS,n}(\phi,\psi)-C\right]^{2}d\phi d\psi\right)^{\frac{1}{2}}
\]

where $\Gamma$ is the region in dihedral space such that $F_{REF}(\phi,\psi)-\min(F_{REF}(\phi,\psi))<25$
kJ/mol and A is the corresponding area. Note that $\Gamma$ includes
all the minima as well as the transition states. The value of $C$
is chosen so as to align the averages of $F_{REF}$ and $F_{MTS}$
over $\Gamma$. As shown in Fig. \ref{fig:fig4}A, both $|\Delta\Delta F|$
and $\epsilon$ slowly grow with $n$. Notably, for $n\leq 8$, the error
on the free energy estimate is comparable with the statistical fluctuations
obtained with $n=1$. The sampling error introduced by the multiple
time step can be effectively monitored by considering the bias effective
energy drift (Fig. \ref{fig:fig4}B). This is particularly convenient in real-case
scenarios, where one does not have a reference simulation to compare
with.

\subsection{Electrostatic energy in a protein/RNA complex}
As a third and more realistic test case we consider the calculation
of the free-energy landscape using metadynamics on a protein/RNA complex. The
nonstructural protein 3 from the Hepatitis C Virus (NS3 HCV) is a motor
protein that unwinds and translocates the double and single stranded
nucleic acids in the 3' $\rightarrow$ 5' direction hydrolyzing ATP
\cite{pyle-helicases}. Here we compute the free-energy profile of
the NS3 helicase complexed with a single stranded RNA, performing a short well-tempered
metadynamics. We use here as a collective variable the Debye-H\"uckel
approximation to the electrostatic component of protein/RNA interaction defined as
in Ref.~\cite{do+13jctc}
\begin{equation}
G^{DH}=\frac{1}{k_{B}T\epsilon_{w}}\sum_{i\in\text{prot}} \sum_{j\in\text{RNA}}q_{i}q_{j}\frac{e^{-\kappa|\mathbf{r}_{ij}|}}{|\mathbf{r}_{ij}|}\label{eq:g-dh}
\end{equation}
where $\epsilon_{w}$ is the water dielectric constant, $q_{i}$ is
the charge of atom $i$, ${\bf r}_{ij}$ is the vector connecting
atoms $i$ and $j$ and $1/\kappa$ is the screening length. Here
atom indexes $i$ and $j$ run over the protein and RNA respectively.
Although we do not characterize protein/RNA binding in this application,
this is a prototype setup for the study of complexes of
charged molecules (e.g. nucleic acids, charged ligands, etc). All
the simulation parameters were the same as for alanine dipeptide,
but simulations were performed in the NPT ensemble using a Parrinello-Rahman
barostat with an isotropic pressure coupling of 1 bar \cite{pr-barostat}
and the AMBER99sb-ILDN{*} force field \cite{hummer_amber,ildn_amber} with parmbsc0 and $\chi_{OL3}$ corrections \cite{pere+07bj,zgarbov2011}.
The system was prepared starting from the crystal structure of the complex between the protein and an ssRNA of 6 nucleotides by Appleby et al. \cite{Appleby} (PDB: 3O8C). The simulation box contains
100012 atoms, and includes 31058 water molecules and NaCl 0.1 M (70
Na\textsuperscript{+} and 62 Cl\textsuperscript{-} ions).

\begin{figure}
\includegraphics[clip,scale=0.3]{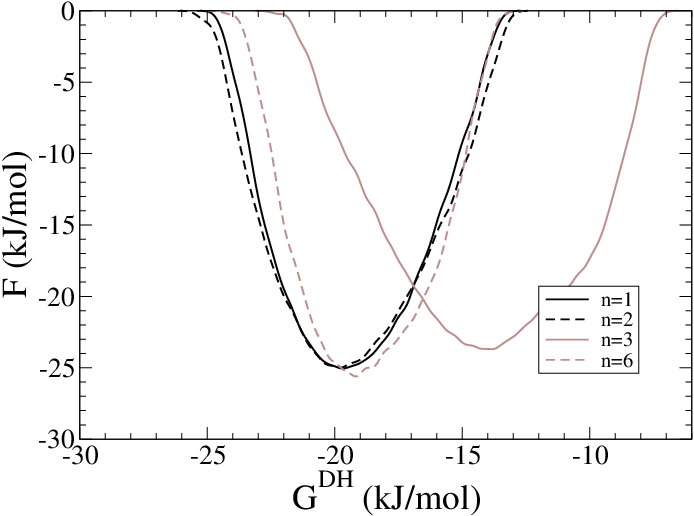}
\caption{\label{fig:FES-NS3}Free-energy profile as a function of the $G^{DH}$
collective variable (see Eq. \ref{eq:g-dh}) as obtained from a well-tempered
metadynamics performed on the protein/RNA complex. Results from different values of the MTS order $n$ are shown.}
\end{figure}

The biased
CV ($G^{DH}$) was computed using a nominal ionic strength of 0.1M.
Protein and RNA here contain respectively $\approx 6500$ and $\approx 180$ atoms,
so that the cost of evaluating the CV is relevant.
Bias was deposited with a Gaussian width of 0.25 kJ/mol, a bias factor
of 2, an initial Gaussian height of 0.1 kJ/mol and a deposition pace
of 0.12 ps. The low value of the bias factor was chosen to avoid the 
complex to dissociate resulting in a difficult-to-converge free-energy landscape.
We performed well-tempered metadynamics simulations of
3 ns length testing different values for the multiple time step order
($n=$1,2,3,6). We notice that it is very difficult to obtain a converged
free-energy landscape with these very short simulations. However,
as it can be seen in Fig. \ref{fig:FES-NS3}, the free-energy landscape
obtained after 3 ns is very consistent across simulations performed
with values of $n$=1 and 2. On the contrary, with $n=3$ we obtain
a clear systematic error in the computed landscape. Remarkably,
the calculation with $n=6$ leads again to a relatively small error.
The particularly bad case of $n=3$ could be related to resonance effects
between the biased CV and the internal degrees of freedom of the system.

Also in this case one can compute the bias effective energy drift for different
choices of $n$. It is worth noting that two extra contributions have
to be included here in order to compute the effective energy properly,
 one to take into account the out of equilibrium nature of metadynamics (see Appendix A) and another one to take into account the change in the cell volume due to the
barostat at every time step (see Appendix B). As it can be appreciated in Fig. \ref{fig:drift-NS3},
the drift is clearly increasing with $n$.
Already at $n=3$ a drift of several tens of kJ/mol/ns can be observed. We notice that electrostatic interaction
is not a smooth function of the atomic coordinates. Indeed, in standard MD codes, only its long range tail
is sometime integrated using a larger time step. Thus, a large drift for $n=3$ can be expected here.
Notably, the drift at $n=2$ is much smaller, consistently with the fact that no
systematic error is observed in Fig.~\ref{fig:FES-NS3}.
This suggest that effective energy conservation is a sufficient criterion to
assess the integration with our MTS scheme.

This application can be
considered as an example of a realistic system. Performance that can
be obtained with this setup are shown in Figure \ref{tab:Performance}.
As it can be seen for a highly expensive CV the performance gain can
be large. Notably, the contribution to the total calculation time 
of the calculation of the CVs scales inversely with the MTS order $n$.
Even for $n=2$ the gain is significant.

\begin{figure}
\begin{centering}
\includegraphics[clip,scale=0.3]{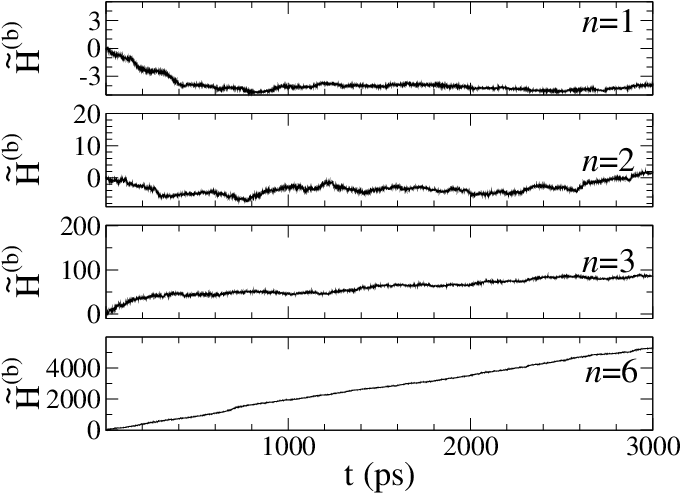}
\par\end{centering}

\caption{\label{fig:drift-NS3}Bias effective energy as a function of time
as obtained from a well-tempered metadynamics simulation biasing the
$G^{DH}$ collective variable performed on the protein/RNA complex.
Results for different values
of the MTS order $n$ are shown as indicated. Energies are in kJ/mol.}
\end{figure}

\begin{figure}
\includegraphics[clip,scale=0.3]{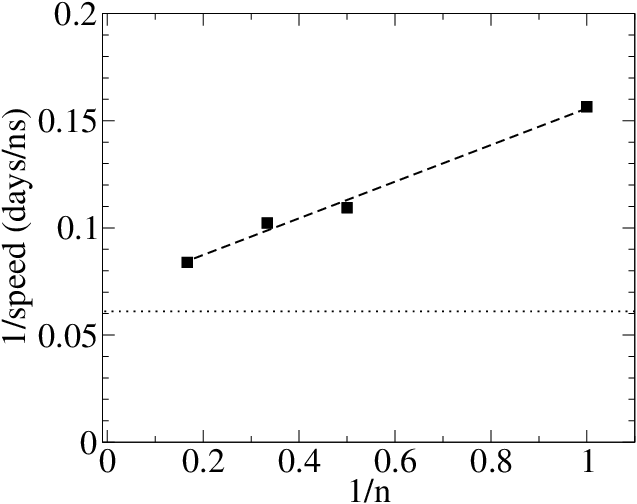}
\caption{\label{tab:Performance}Performance measured on the metadynamics calculation
on a protein/RNA complex for different multiple time step orders $n$ (squares),
estimated on a GPU-based Linux workstation (CPU: 12 cores Intel E5-2620;
GPU: 1 GeForce GTX TITAN). A linear fit is also shown (dashed line).
Both force-field calculation and CV calculation
were parallelized on 12 threads using OpenMP.
Performance without metadynamics is indicated as a horizontal dotted line .}
\end{figure}

\section{Discussion and Conclusions}
We presented a scheme based on reversible multiple time step to integrate
physical forces and biasing forces using different strides. This can
be convenient when biasing collective variables (CVs) that are smooth and computationally
expensive, since it allows the CVs to be computed every few steps.
The implementation is straightforward, since it just requires biasing
forces to be scaled by a factor $n$ equal to the stride used for
their calculation.
The accuracy of the algorithm is discussed in detail on a one dimensional model
and on alanine dipeptide in explicit solvent. Additionally,
a protein/RNA complex has been used as a model for a realistic
application. Although in this case it is difficult to converge
a free-energy landscape within a short simulation,
the impact of the introduced algorithm on performance is clear.
A multiple time step order $n=2$ has been shown to provide a remarkable speed up
without significantly affecting the accuracy of the calculation.
We stress that the case of electrostatic interaction is particularly difficult
since this CV is not a smooth function of the atomic coordinates.
In the alanine dipeptide, where CVs that are smooth functions of the coordinates were
employed,
values up to $n\approx 8$ where shown to introduce no significant error.
In practical applications, different CVs could require different values
of $n$ and thus the performance gain could be different.
To help the users in this choice, we introduced here a 
theoretical framework for assessing the accuracy of the explored
ensemble.
 This formalism is very powerful in that it allows an energy
drift to be defined that includes selected portions of the potential
energy. We here used it to isolate the contribution to the effective energy
drift given by the biasing force from the contribution given by the
physical forces. This decomposition allows one to focus the assessment
to the relevant portion of the potential-energy function, thus avoiding
relevant errors to be masked by e.g. solvent fluctuations.
In this manner, although sampling errors are not corrected, it is easy
to detect improper choice of $n$.
 Even
though it is not possible to define a universal relationship between
the drift and the error, it is clear that the effective energy is
a powerful tool to validate MTS simulations.
Finally, we observe that
this check can in principle be used also with a single time step integration
scheme ($n=1$) since, also in this case, the forces added by enhanced
sampling techniques could be too stiff to be integrated with a conventional
time step.

The discussed algorithm has been implemented in a locally modified
version of the PLUMED plugin and is available on request.  It will be publicly
available in the next PLUMED release.

\section{Appendix A: Out of equilibrium simulations}
The effective energy drift provides a rigorous way to asses the validity
of detailed balance. However, there are a number of relevant situations
where detailed balance is by construction violated during the simulation,
e.g. all non equilibrium free-energy techniques \cite{dellago2013computing}.
As an example, in steered molecular dynamics \cite{grubmuller1996ligand}
the work can be used to compute free energies \cite{jarzynski1997nonequilibrium}
or to reweight the obtained conformations \cite{coli-buss12jacs}.
However, for the sake of validating the integration of the equations
of motion, one can limit the analysis to the portion of work performed
by the integrator. In case of steered molecular dynamics the work
accumulated by moving the restraint should be removed from the effective
energy drift. In a similar fashion, in metadynamics \cite{laio-parr02pnas}
the sum of the height of the accumulated Gaussians should be subtracted
from the effective energy drift.

\section{Appendix B: Variable cell simulations}
In the case of variable cell simulations \cite{ande80jcp,pr-barostat}
an additional contribution related to cell change should be added
to the effective energy drift. Additionally, attention should be paid
when computing the increment of positions $\Delta\mathbf{q}$ when
particles cross the boundaries. Equation \ref{eq:bias-drift} should
be replaced with

\begin{equation}
\Delta\tilde{H}^{(b)}(t)=\sum_{i=1}^{N_{at}}\Delta\mathbf{s}_{i}\cdot\left(\frac{\mathbf{g}_{i}^{(b)}(t)+\mathbf{g}_{i}^{(b)}(t+\Delta t)}{2}\right)+\mathrm{tr}\left[\Delta\mathbf{h}^{T}\left(\frac{\mathbf{W}^{(b)}(t)+\mathbf{W}^{(b)}(t+\Delta t)}{2}\right)\right]+\Delta V\label{eq:npt}
\end{equation}

where $\mathbf{h}$ is the matrix of lattice vectors (as rows), $\mathbf{s}_{i}=\left(\mathbf{h}^{T}\right)^{-1}\mathbf{q}_{i}$
are the scaled coordinates of $i$-th atom,  $\mathbf{g}_{i}=\mathbf{h}\mathbf{f}_{i}$
are the scaled biasing forces on $i$-th atom, and $\mathbf{W}^{(b)}=\frac{\partial V}{\partial\mathbf{h}}$
is related to the bias contribution to the virial tensor. The first
term in Eq. \ref{eq:npt} is defined in terms of scaled coordinates
and forces so as to allow the distance vector $\Delta\mathbf{s}_{i}$
to be evaluated with the minimal image convention when the cell is
modified. This term reduces to the first term in Eq. \ref{eq:bias-drift}
when the cell matrix is constant. The second term is only present
for variable cell simulations.

\section{Acknowledgement}
Michele Ceriotti is acknowledged for carefully reading the manuscript and providing useful suggestions.
Alessandro Laio is also acknowledged for useful discussions.
The research leading to these results has received funding from the
European Research Council under the European Union\textquoteright{}s
Seventh Framework Programme (FP/2007-2013) / ERC Grant Agreement n.
306662, S-RNA-S.

\section{Supplementary Information}

Figures showing the dependence of the bias effective energy drift on the friction coefficient $\gamma$ for the 1D model.
Figure showing the effective energy drift for the 1D model using a Nos\'e-Hoover thermostat.


\providecommand{\latin}[1]{#1}
\providecommand*\mcitethebibliography{\thebibliography}
\csname @ifundefined\endcsname{endmcitethebibliography}
  {\let\endmcitethebibliography\endthebibliography}{}

\end{document}